\documentclass[journal]{IEEEtran}
\usepackage{graphicx}
\usepackage{textcomp}
\usepackage{nth}

\begin{document}
\title{FACT -- The G-APD revolution\\in Cherenkov astronomy}

\newcommand{\ethz}{$^1$}
\newcommand{\tudo}{$^2$}
\newcommand{\unige}{$^3$}
\newcommand{\uniw}{$^4$}
\newcommand{\epfl}{$^5$}

\newcommand{\uniz}{$^{a}$}
\newcommand{\kynu}{$^{b}$}
\newcommand{\mpim}{$^{c}$}
\newcommand{\tum}{$^{d}$}

\author{
T.~Bretz\ethz$^{,*}$,\\
H.~Anderhub\ethz,
M.~Backes\tudo,
A.~Biland\ethz,
V.~Boccone\unige,
I.~Braun\ethz,
J.~Bu\ss\tudo,
F.~Cadoux\unige,
V.~Commichau\ethz,
L.~Djambazov\ethz,
D.~Dorner\uniw,
S.~Einecke\tudo,
D.~Eisenacher\uniw,
A.~Gendotti\ethz,
O.~Grimm\ethz,
H.~von Gunten\ethz,
C.~Haller\ethz,
C.~Hempfling\uniw,
D.~Hildebrand\ethz,
U.~Horisberger\ethz,
B.~Huber\ethz\uniz,
K.-S.~Kim\ethz\kynu,
M.~L.~Knoetig\ethz,
J.-H.~K\"ohne\tudo,
T.~Kr\"ahenb\"uhl\ethz,
B.~Krumm\tudo,
M.~Lee\ethz\kynu,
E.~Lorenz\ethz\mpim,
W.~Lustermann\ethz,
E.~Lyard\unige,
K.~Mannheim\uniw,
M.~Meharga\unige,
K.~Meier\uniw,
S.~M\"uller\tudo,
T.~Montaruli\unige,
D.~Neise\tudo,
F.~Nessi-Tedaldi\ethz,
A.-K.~Overkemping\tudo,
A.~Paravac\uniw,
F.~Pauss\ethz,
D.~Renker\ethz\tum,
W.~Rhode\tudo,
M.~Ribordy\epfl,
U.~R\"oser\ethz,
J.-P.~Stucki\ethz,
J.~Schneider\ethz,
T.~Steinbring\uniw,
F.~Temme\tudo,
J.~Thaele\tudo,
S.~Tobler\ethz,
G.~Viertel\ethz,
P.~Vogler\ethz,
R.~Walter\unige,
K.~Warda\tudo,
Q.~Weitzel\ethz,
M.~Z\"anglein\uniw\\  
(FACT Collaboration)\\[1ex]
$^{*}${\em Corresponding author: Thomas Bretz (tbretz@phys.ethz.ch)}\\[1.5em]
\thanks{\ethz ETH Zurich, Switzerland --
   Institute for Particle Physics, Schafmattstr.~20, 8093 Zurich}
\thanks{\tudo Technische Universit\"at Dortmund, Germany --
   Experimental Physics 5, Otto-Hahn-Str.~4, 44221 Dortmund}
\thanks{\unige University of Geneva, Switzerland --
   ISDC, Chemin d'Ecogia~16, 1290 Versoix --
   DPNC, Quai Ernest-Ansermet 24, 1211 Geneva}
\thanks{\uniw Universit\"at W\"urzburg, Germany --
   Institute for Theoretical Physics and Astrophysics,
   Emil-Fischer-Str.~31, 97074 W\"urzburg}
\thanks{\epfl EPFL, Switzerland --
   Laboratory for High Energy Physics, 1015 Lausanne}
\thanks{\uniz Also at: University of Zurich, Physik-Institut, 8057 Zurich,
   Switzerland}
\thanks{\kynu Also at: Kyungpook National University, Center for High Energy
   Physics, 702-701 Daegu, Korea}
\thanks{\mpim Also at: Max-Planck-Institut f\"ur Physik, 80805 Munich,
   Germany}
\thanks{\tum Also at: Technische Universit\"at M\"unchen, 85748 Garching,
   Germany}
\thanks{}
\thanks{The important contributions from ETH Zurich grants ETH-10.08-2 and
ETH-27.12-1 as well as the funding by the German BMBF (Verbundforschung
Astro- und Astroteilchenphysik) are gratefully acknowledged. We thank
the Instituto de Astrofisica de Canarias allowing us to operate the
telescope at the Observatorio Roque de los Muchachos in La Palma, the
Max-Planck-Institut f\"ur Physik for providing us with the mount of the
former HEGRA CT\,3 telescope, and the MAGIC collaboration for their
support. We also thank the group of Marinella Tose from the College of
Engineering and Technology at Western Mindanao State University,
Philippines, for providing us with the scheduling web-interface.}
}

\maketitle
\pagestyle{empty}
\thispagestyle{empty}

\begin{abstract}
Since two years, the FACT telescope is operating on the Canary Island
of La Palma. Apart from its purpose to serve as a monitoring facility
for the brightest TeV blazars, it was built as a major step to
establish solid state photon counters as detectors in Cherenkov
astronomy. The camera of the First G-APD Cherenkov Telesope comprises
1440 Geiger-mode avalanche photo diodes (G-APD), equipped with solid
light guides to increase the effective light collection area of each
sensor. Since no sense-line is available, a special challenge is to
keep the applied voltage stable although the current drawn by the G-APD
depends on the flux of night-sky background photons significantly
varying with ambient light conditions. Methods have been developed to
keep the temperature and voltage dependent response of the G-APDs
stable during operation. As a cross-check, dark count spectra with high
statistics have been taken under different environmental conditions. In
this presentation, the project, the developed methods and the
experience from two years of operation of the first G-APD based camera
in Cherenkov astronomy under changing environmental conditions will be
presented. 
\end{abstract}


\section{Introduction}
%
%
%
%


\begin{figure}[!htb]
\centering
\includegraphics[width=0.49\textwidth]{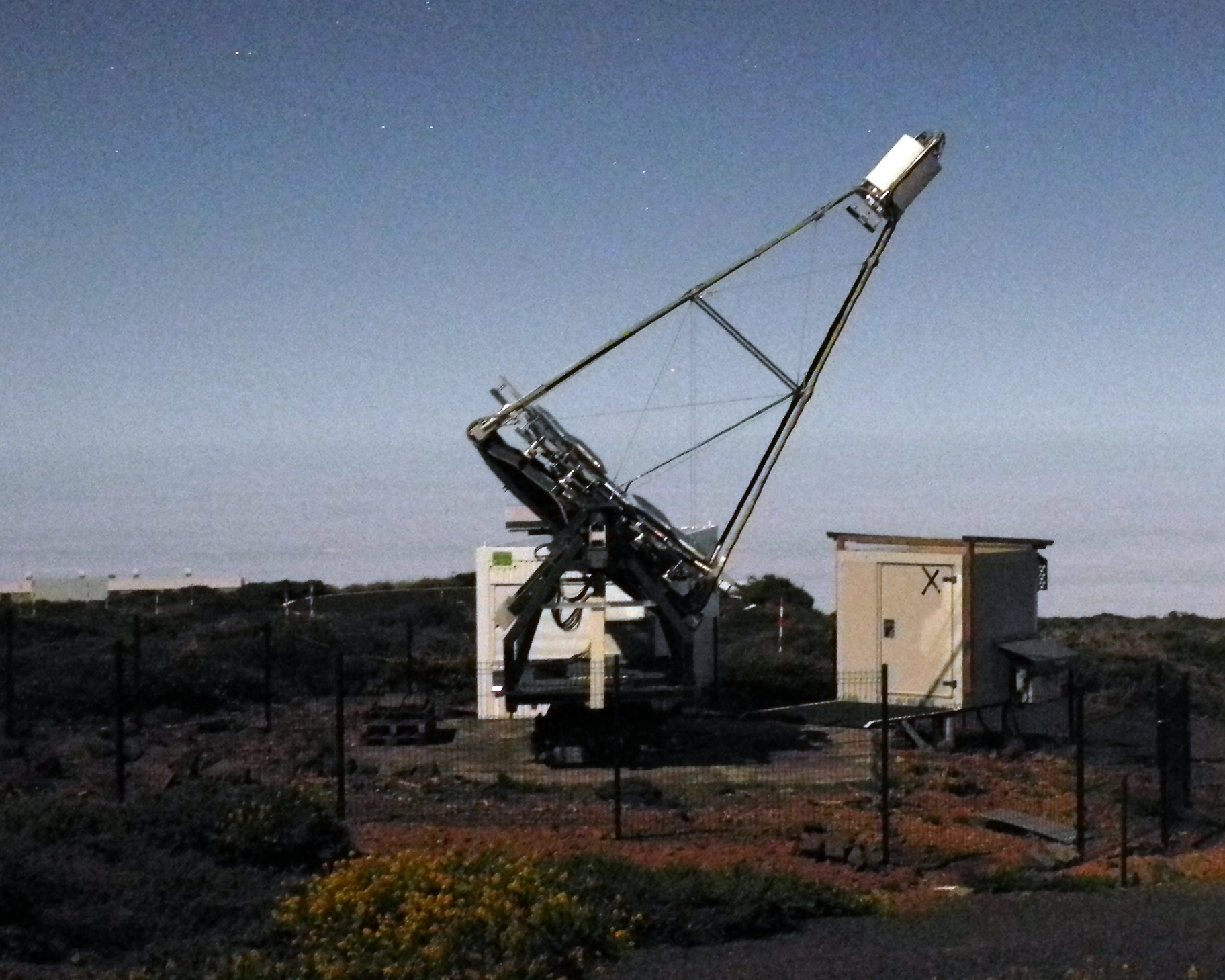}
\caption{Picture of the FACT telescope during observations at night. In the background, the white container hosting the computing system, the bias power sully and the drive system is visible. Courtesy of D.~Dorner.}
\label{fig:pic}
\end{figure}

For the first time in Cherenkov astronomy, photo multiplier tubes (PMT)
for photo detection have been replaced by more modern silicon based
photo sensors. The First G-APD Cherenkov Telescope (FACT) has
implemented a focal plane  using 1440 Hamamatsu MPPC S10362-33-050C
sensors for photo detection. A picture of the telescope can be
seen in Fig.~\ref{fig:pic}. A detailed description of its hardware and
software is given in the FACT design report \cite{bib:design}.

Apart from the use of recent technology for photo detection, the FACT
project aims towards long-term monitoring of the brightest known blazars at
TeV energies. Due to the natural gap in observations, which arises from
bright moon light when operation with photo multiplier tubes is hardly
possible with standard settings, and the busy
schedule of existing high performance instruments, the existing
sampling density is comparably low in the order of a few minutes every
few days. Therefore, the aims is a high duty cycle instrument
with the best possible data taking efficiency and observations during
strong moon light, so that several hours per night and source can be achieved
and the gap around full moon can be filled.

Data taking efficiency and the results of almost two years of
successful monitoring of Markarian 421 and Markarian 501 are presented
at the end of the paper. In the following, the use of Geiger-mode
avalanche photo sensors in the camera are discussed in more details and
results from stability measurements are presented.

\subsection{Application of G-APDs}

In Cherenkov astronomy, photo sensors are used to detect Cherenkov
light emitted by particle cascades in the atmosphere. The image of
these particle showers is then used to reconstruct the properties of
the primary particles such as particle type, energy and origin.

To detect these nano-second short light-flashes consisting of a few
tens to hundreds of photons, sensitive and fast photo sensors are
needed. Since its beginning, Cherenkov astronomy has used photo
multiplier tubes for photo detection, but for several years silicon
based photo sensors are on the market as well. During the past years,
Geiger-mode avalanche photo diodes (G-APD) became powerful and inexpensive enough to
replace photo multiplier tubes in Cherenkov astronomy.

Each sensor applied in the FACT camera consists of 3600 single 
Geiger-mode avalanche photo diodes (cells),
50\,\(\mu\)m\,x\,50\,\(\mu\)m each. Since the cells are operated above
their breakdown voltage, the so-called Geiger-mode, each detected
photon triggers a complete discharge of a single cell. The resulting
pulse is comparable in amplitude and shape for all cells. Due to the
high precision of the production, the fluctuations from cell to cell
are small. 

Devices currently on the market achieve photo detection efficiencies
comparable to the best available PMTs, and improved photo detection
efficiency is expected for the next generation. 

As compared to PMTs, silicon based photo sensors are easier to handle
due to their lower bias voltage in the order of 100\,V or below, they
are mechanically more robust, and they are insensitive to magnetic
fields. Apart from their high future potential, their main advantage in
Cherenkov astronomy is their robustness against bright light which
allows operation under bright moon light conditions.

As a drawback, G-APDs encounter a high dark count rate, so-called
optical crosstalk and a comparably high afterpulse probability. 

The dark count rate of the applied sensors is in the order of 1\,MHz to
9\,MHz for typical operation temperatures between 0\,\textdegree{}C to 
30\,\textdegree{}C. At the same time, the rate of detected photons from
the diffuse night sky background is in the order of 30\,MHz to 50\,MHz
even during the darkest nights. Consequently, the dark count rate of
the sensor is negligible.

\subsection{Dark count rate, optical crosstalk and afterpulses}
\begin{figure*}[!htb]
\centering
\includegraphics[width=0.48\textwidth]{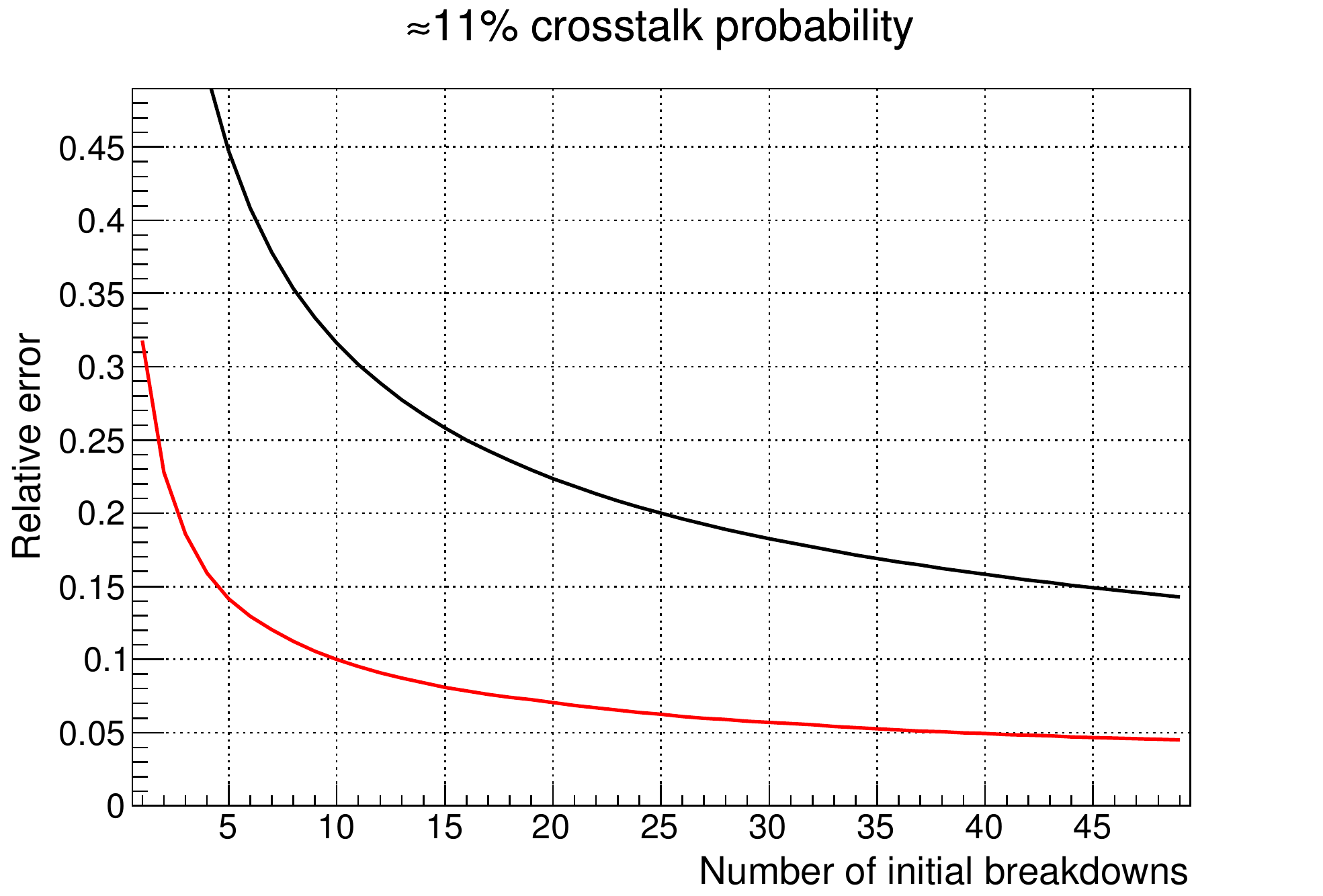}\hfill
\includegraphics[width=0.48\textwidth]{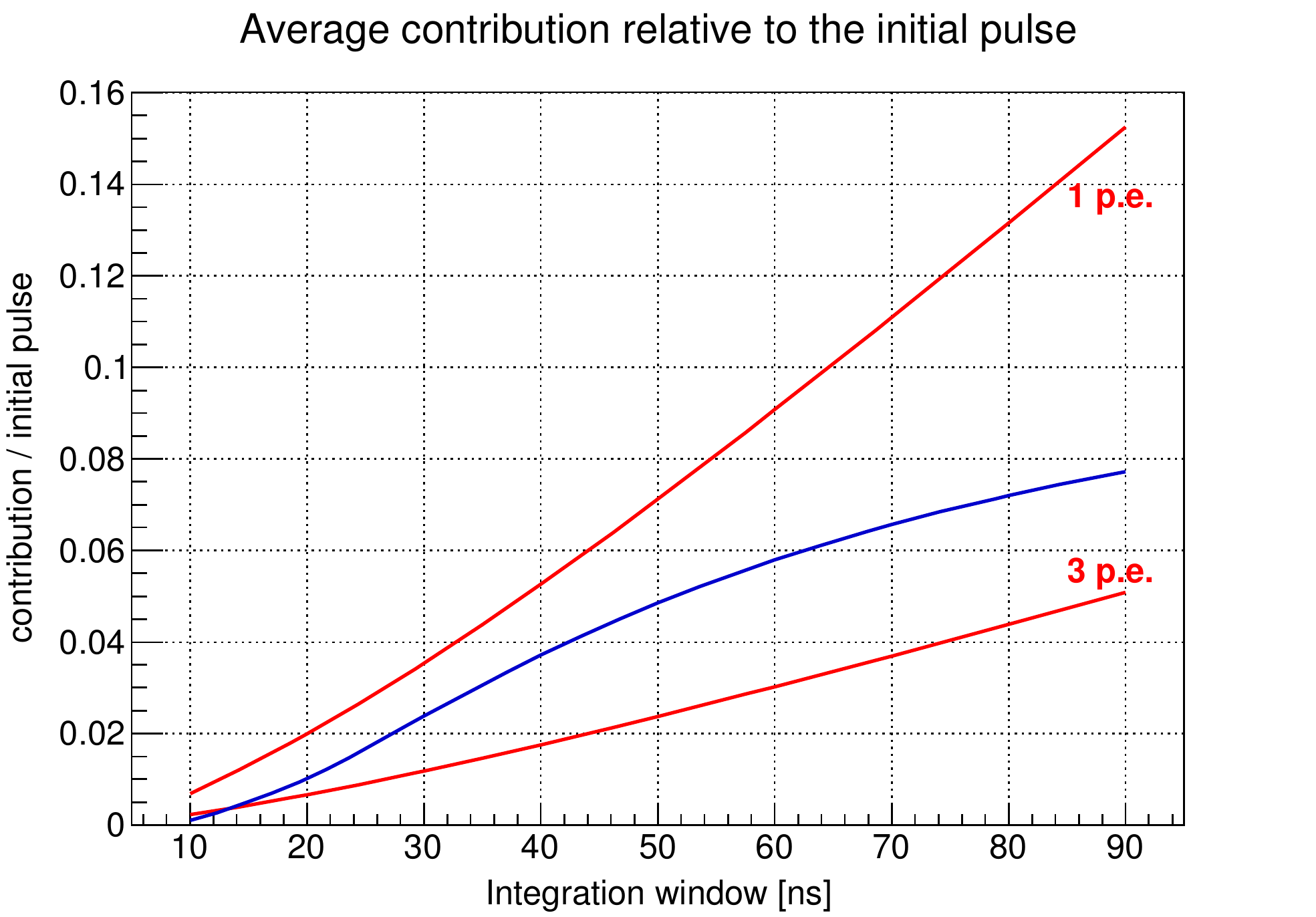}
\caption{Left: The figure shows the relative error introduced on average 
by optical crosstalk (red) and its statistical error (black)
on the signal as a function of the number of initial breakdowns considering
a crosstalk probability of 11\%. In both cases, the error
on a single event can be significantly larger, but on average the statistical
error dominates considerably. Right: The figure shows the additional contribution
from afterpulses (blue) as a fraction of the unbiased pulse as a function of the applied
pulse integration window starting pulse integration at the half height leading edge
compared to the contribution from optical crosstalk (red). While the relative
contribution by afterpulses does not depend on the initial number of breakdowns,
the relative contribution from optical crosstalk decreases with an increasing number of
breakdowns. While the contribution from afterpulses in single events can reach 100\%
the number of biased events is so small, that the average contribution for integration
windows of less than 40\,ns is below 4\%.}
\label{fig:result}
\end{figure*}

The so-called optical crosstalk is the probability that the discharge of a
G-APD cell will indirectly trigger at least one other cell due to
photons potentially emitted during the breakdown. While this effect is
significant for signals in the order of one, for larger signals it just
increases the amplitude and its fluctuation statistically. 
Figure~\ref{fig:result} shows the relative error introduced from crosstalk
compared to the statistical error. Since the signals of Cherenkov
showers are comprised of a number of photons large compared to one, and
have an intrinsic fluctuation of the order of a few percent, optical
crosstalk needs to be considered quantitatively in the shower
reconstruction, but has no significant effect on the quality.

Afterpulses in G-APDs originate from trapped charges released with a
short delay after the initial pulse. Their probability is 
exponentially decreasing with a time scale in the order of the decay
time of the pulse, see \cite{bib:afterpulses}. Figure~\ref{fig:result} shows the 
influence of afterpulses as a function of the integration window.
Although their total probability in the order of 10\% to 20\% is high
compared with PMTs, they usually occur within the timescale of the
initial pulse. The exponential decrease of the probability ensures
that afterpulses induced by coincident signals in several cells or
sensors are not released synchronously. Therefore, afterpulses in
G-APDs can not fake trigger signals in contrast to afterpulses in
PMTs. In addition, due to the partial discharge of the cell after the
initial pulse, the amplitude of early afterpulses is highly decreased.
Generally, their absolute amplitude, i.e. falling edge plus afterpulse,
does not exceed the amplitude of the initial pulse. In the pulse
extraction algorithm, their influence can be suppressed strongly, if the
integration range is limited to a short interval around the peak of the
initial pulse.

\subsection{Breakdown voltage}

The main challenge in the application of G-APDs in an experiment
exposed to changing environmental conditions is the dependence of their
breakdown voltage from the temperature and the voltage drop induced by 
high currents. 

Cherenkov telescopes are exposed to natural temperature changes
during the night. Since the breakdown voltage of the G-APDs in use is temperature 
dependent (\(\approx\)55\,mV/K), the gain will follow the temperature,
if the voltage is not adjusted accordingly.

In addition, each sensor has a network of serial resistors. When the moon rises
during the night, more breakdowns take place due to the increased
photon flux and induce a higher current in the resistors. The
voltage drop induced at the serial resistors is enough to significantly
decrease the gain of the sensor. Therefore, this voltage drop must be
determined and the applied voltage has to be adjusted accordingly.

In the following, the calibration and adjustment procedure applied
in the FACT camera is presented and discussed.

\section{Bias voltage calibration}

The camera contains a total of 1440 photo sensors and 320 bias voltage 
channels. Four respectively five sensors are combined into a
single bias channel at a time.
To avoid influences of different operation voltages of the sensors combined into 
a single channel, the sensors in each channel are ordered together
according to the operation voltage provided by the manufacturer.
The operation voltage is provided with a precision of 0.01\,V which corresponds
to a precision of \(\approx\)1\% in gain at the operation voltage.

\begin{figure*}[!htb]
\centering
\includegraphics[angle=90,height=\textheight]{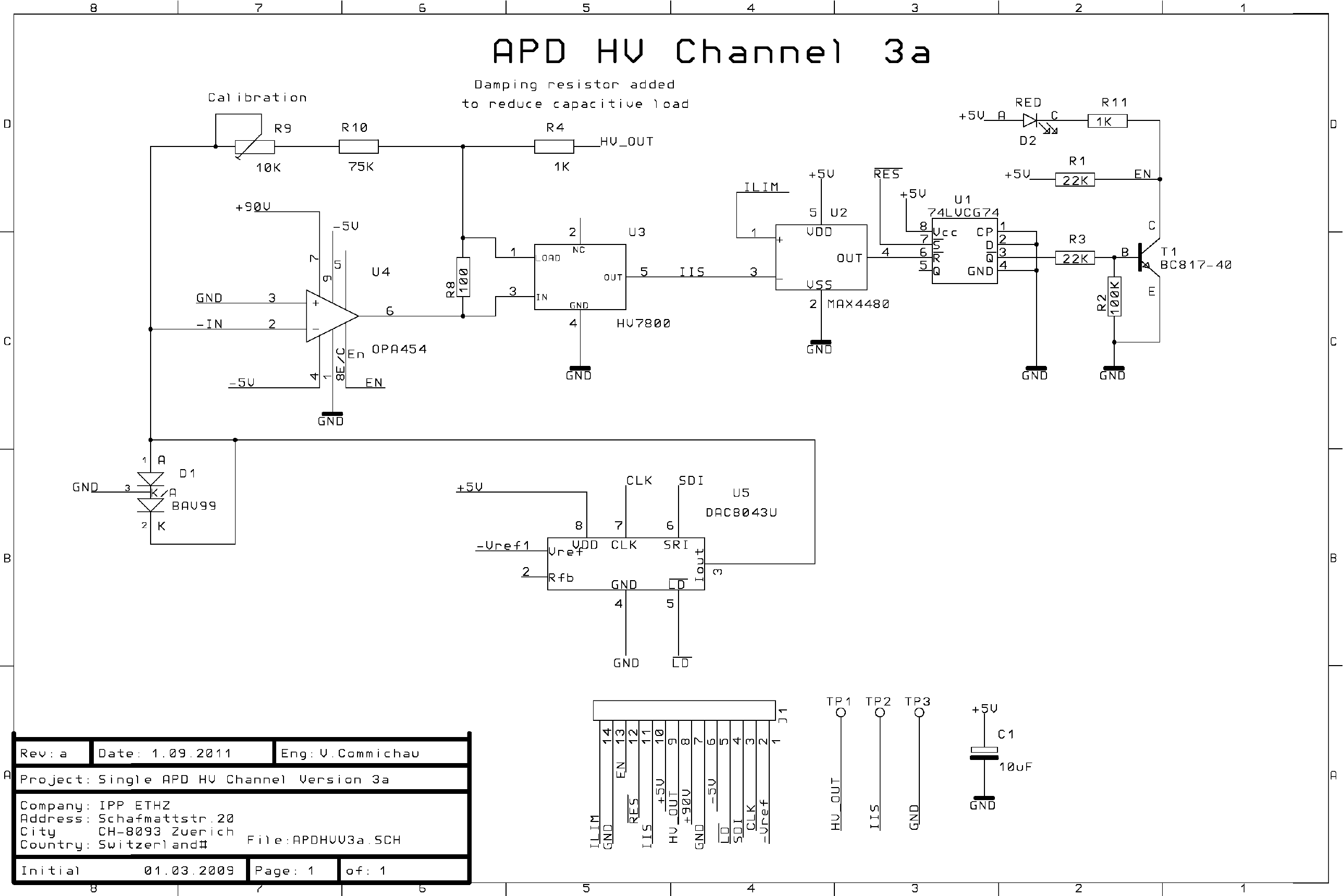}
\caption{Schematics of the circuit which provides the bias power for the
G-APDs. It is controlled by an digital-to-analog converter and the drawn current
is measured by an analog-to-digital converter.}
\label{fig:schematics}
\end{figure*}

The schematics for a reference channel is shown in
Fig.~\ref{fig:schematics} at the end of the paper. The circuit is driven by a current supplying
12\,bit digital-to-analog converter (DAC) with a maximum current output
of 1\,mA. Its output voltage is amplified by an op-amp (OPA\,454) which
is operated in a programmable voltage source configuration, c.f.~its
datasheet. A calibration resistor in the feedback loop allows to
adjust the voltage output of each channel. A measurement circuit with a
12\,bit analog-to-digital converter (ADC) measures the current drawn by
both, the feedback loop and the attached sensors. The voltage output
and camera input have damping resistors, 1\,k\(\Omega\) each, to avoid
oscillations by the capacitance of the cable. Each sensor has its own
serial resistor (3.9\,k\(\Omega\)).

With a nominal calibration resistor of 90\,k\(\Omega\), the voltage
output realized by the circuit for the full range of the DAC is between
0\,V and 90\,V in steps of 22\,mV. The current can be measured with a
precision of 1.2\,\(\mu\)A in a range between 0\,A and 5\,mA.

In an ideal circuit with an ideal op-amp, no further calibration is
necessary. A real circuit must be calibrated to correct for the non-ideal
behavior of the op-amp and the limited accuracy with which the total
resistances are known. This is necessary to achieve the maximum possible
precision in the absolute voltage output of 22\,mV at roughly 70\,V corresponding to 
about 0.3\%. To calibrate the circuit, the
voltage output at U\_OUT and the corresponding feedback value from the
ADC are measured as a function of the DAC setting. This measurement
takes place well below the breakdown voltage of the G-APD to ensure
that the sensors do not draw any current through U\_OUT, neither in
avalanche mode nor in Geiger-mode.

\subsection{Correcting the voltage drop}

The calibration of the voltage output versus DAC setting is mainly necessary
due to the non-ideal behavior of the op-amp. Since a non
ideal op-amp usually draws current on its input, the measured current
at the ADC is not exactly identical with the current provided by the
DAC and needs to be calibrated as well.

Above 50\,V, the behavior of the circuit can be considered linear within the
precision of the DAC setting, so that only slope and offset need to be calibrated.


The output voltage \(U_{out}\) at U\_OUT can be expressed as a
function of the DAC setting as 
\[U_{out} = u + r\cdot I_{dac}\]
with the calibration constants \(u\) and \(r\) determined by calibration measurements.

With the ADC, the voltage drop at \(R_8\) is measured precisely, corresponding to the current \(R_8\), and below the breakdown voltage of the G-APDs when no current flows through \(R_4\) also at \(R_9\) and \(R_{10}\). Therefore, it gives the current \(I_9\) as a function of the DAC setting assuming that the imprecisions of \(R_8\) and the
imprecision of the measurement curcuit can be neglected against the deviations of
\(I_9\) from the output current of the DAC:

\[I_9 = i + c\cdot I_{dac}\]
with the calibration constants \(i\) and \(c\) determined by the calibration measurements. 

%
%
%
%
%
%
%
%
%
%
%
%
%
%
%


Below the breakdown voltage, the voltage at the sensors is then equal
to the calibrated output voltage U\_OUT. If the breakdown voltage is 
exceeded, an additional current will flow through U\_OUT and the
100\,\(\Omega\) measurement resistor induces an additional voltage drop
which has to be taken into account.

The current leaking through U\_OUT can then be calculated as the
difference between the current \(I_8\), equal to the measured current
\(I_{adc}\), and the current \(I_9\) through \(R_9\) and \(R_10\):

\[I_{out} = I_8 - I_9 = I_{adc} - I_9 = I_{adc} - (i + c\cdot I_{dac})\]

The voltage drop at the resistors \(R_4\), \(R_5\) and \(R_n\) for \(N\) parallel sensors 
is calculated by

\[U_{drp} = (1\,\mbox{k}\Omega + 1\,\mbox{k}\Omega + \frac{3.9\,\mbox{k}\Omega}{N})\cdot I_{out}.\]

The resistor \(R_5\) is an additional serial resistor of 1\,k\(\Omega\) in
each bias channel on the camera side, while \(R_n\) are resistors of 3.9\,k\(\Omega\)
within each individual sensor branch.

\subsection{Correcting the change of operation temperature}

In addition to the voltage drop, an intrinsic feature of each resistor
network, the change in operation voltage of the G-APDs by temperature
variations, has to be considered. Therefore, 31 temperature sensors were
homogeneously distributed on the sensor plate. Their values are read out
once every 15\,s and inter- respectively extrapolated linearly to get an estimate of
the temperature of each bias patch. The difference between the minimum
and maximum temperature in the camera never exceeds 4\,K. To correct
for the measured temperature, the change on operation voltage \(\Delta
U\) is calculated as 

\[\Delta U = (25\,^\circ\mbox{C} - T)\cdot t\]

for each patch separately at a patch temperature \(T\) and a temperature
coefficient \(t\). How the temperature for each patch is determined is
described in the next section.


The nominal voltage setting \(U_{set}\) is then derived as the
manufacturer given operation voltage \(U_{op}\), the temperature
correction \(\Delta U\) and the voltage drop \(U_{drp}\) at the nominal set-point:

\[U_{set} = U_{op} + \Delta U(T) + U_{drp}(U_{set}).\]

\section{Temperature measurement}

To achieve a reasonable precision, the average temperature of the
sensors supplied by a single bias voltage channel has to be known.
Therefore, the temperatures measured by the temperature sensors
needs to be inter- respectively extrapolated. Since three of them do
not show any signal, 28 are available in total. It was planned to repair
the remaining three sensors, but since their access is very diffcult and
the camera has to be unmounted, the effort has never been carried out because
no significant improvement is expected.

For interpolation, a so-called Delaunay triangulation is calculated. A
Delaunay triangulation connects each set of three points from a set of
points such that no other point is inside the circumcircle of the three
points. Delaunay triangulations maximize the minimum angle of all the
angles of the triangles in the triangulation and therefore avoid skinny
triangles. 

\begin{figure}[!htb]
\centering
\includegraphics[trim=4cm 0cm 3.5cm 0cm,clip=true,width=0.5\textwidth]{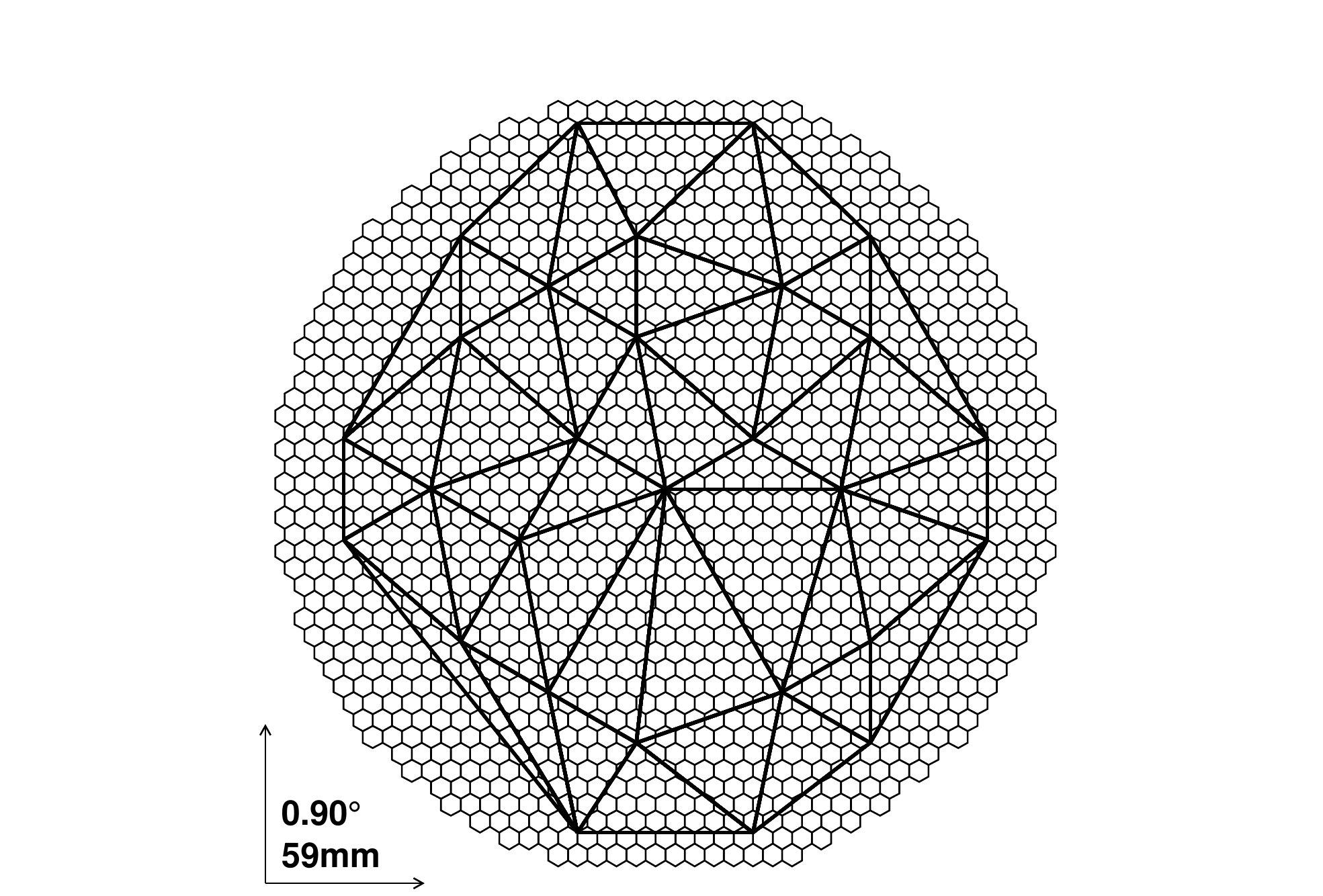}
\caption{This figure shows the Delaunay triangulation for the working
temperature sensors in the camera. Each intersection point corresponds to one
sensor.} 
\label{fig:delaunay} 
\end{figure}

The Delaunay-triangulation for the set of the 28 working sensors has been
calculated. The triangulation is shown in Figure~\ref{fig:delaunay}
where the cross points of the edges of the triangles correspond to the
sensor positions. 

For each point inside any of the triangles,
the temperature can now be interpolated linearly. For each point outside 
the border of the Delaunay triangulation, an extrapolation algorithm is applied.
To get a reasonable and continuous extrapolations, the following algorithm is
applied: If the point does not lay within a triangle, choose the set of
three points for which it lays within their circumcirle. If no such
circle is available, choose the triangle with the closest circle
center. The extrapolation is then done linear for the set of three
points. It should be mentioned, that this algorithm gives comparable
good results very close to the border of the Delaunay-triangulation, but
is not suited for points further away as any extrapolation algorithm.

In principle, more complex interpolation algorithms, e.g. natural
neighbor interpolation or dimensional splines, could be applied.
While the maximum temperature difference in the camera might reach
values of 4\,\textdegree{} just after sunset, the typical maximum
temperature difference in the camera is not more than
1.5\,\textdegree{}C. Even for the most extreme cases, this yields a maximum
temperature difference of 1\,\textdegree{}C in a single triangle
corresponding to 5\% in gain. Therefore, a linear interpolation can be
considered precise enough.


\section{Results}

To measure the stability of the system, two main methods are available.
The first method extracts the gain from the dark count spectrum, the
second uses an external light pulser as signal source. The advantage
using the dark count spectrum, is the availability of a very precise
and unbiased method to measure the gain directly. Evaluating a spectrum
of random signals induced from the diffuse night-sky background under
dark night condition is possible as well, but leads to less stable
results. Extracting single pulses from data taken with additional
background light is extremely chellenge. This limitation can be overcome using a
stable external light source which produces an amplitude well above
the night-sky light level. In contrast to the dark count spectrum,
this method is biased by the transmission of the light-collecting cones,
the photo detection efficiency of the sensors and the stability of the
light source itself. In this context, the dark count spectrum is well
suited to extract the dependency of the gain on the sensor temperature,
while light-pulser measurements provide information about the
dependency on the background light-level or the resulting current.

A third method is to use the physics triggered data itself, so called ratescans,
for a qualitative estimation of the stability. Details can be found in~\cite{bib:ratescans}.

More about that can be found as well in~\cite{bib:stability}.

\subsection{Temperature dependence}

\begin{figure*}[!htb]
\centering
\includegraphics[height=0.29\textwidth]{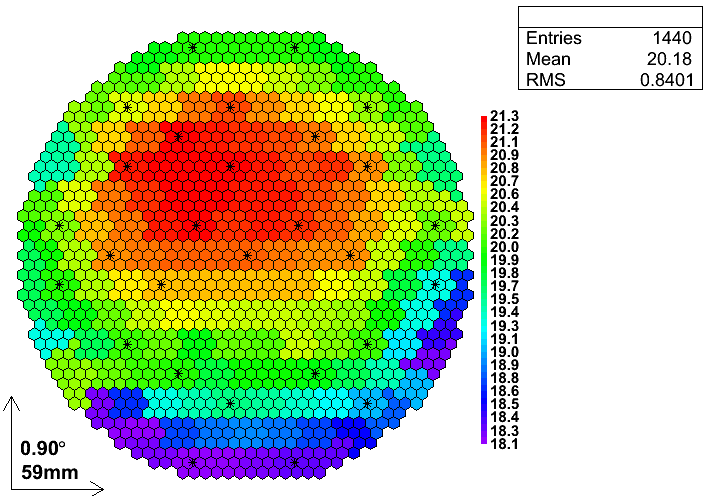}
\includegraphics[height=0.29\textwidth]{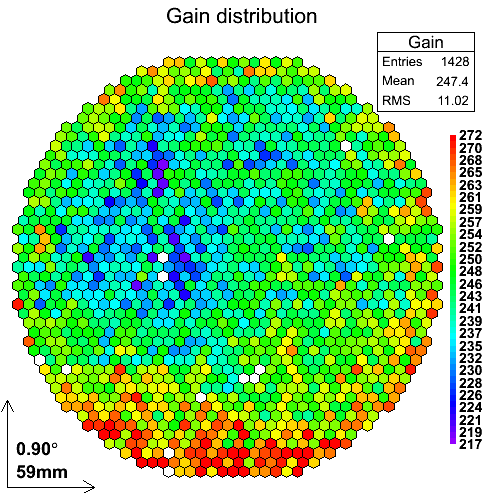}\hfill
\includegraphics[height=0.29\textwidth]{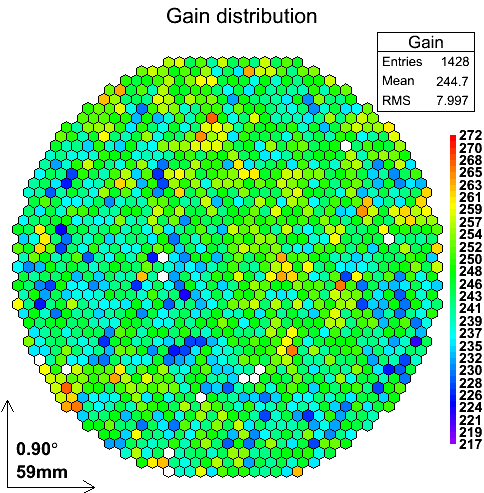}\hfill
\caption{The left camera display shows interpolated temperatures
for all pixels (the same value for all pixels connected to the 
same bias channel) in degree Celsius for data taken just after sunset.
The central camera display shows the gain in arbitrary units extracted
for each pixel, if the applied voltage of all pixels is corrected
for the average temperature of all 28 sensors instead of the individual 
patch temperature. The variance is 4.5\%. The right camera display shows the
gain extracted, if individual patch-wise corrections for the operation
voltage are applied. The variance has decreased to 3.3\%. Both
displays feature identical color scales.} 
\label{fig:temp} 
\end{figure*}

Figure~\ref{fig:temp} shows a typical temperature distribution,
directly after sunset with a maximum temperature difference of
3.2\,\textdegree{}C, a gain distribution in which for all bias channels
an average correction is applied and a gain distribution in which a per
channel correction is applied. Both measurements took place directly
after each other, and the gain was extracted from a dark count spectrum
taken with closed lid. It can be seen that the distribution of gains
without channel-wise correction depends on the temperatures in the
camera as expected. If the temperature correction is applied per bias
channel, an improvement of the width of the distribution from 4.5\% to
3.3\% is visible. 

\begin{figure*}[!htb]
\centering
\includegraphics[width=0.323\textwidth]{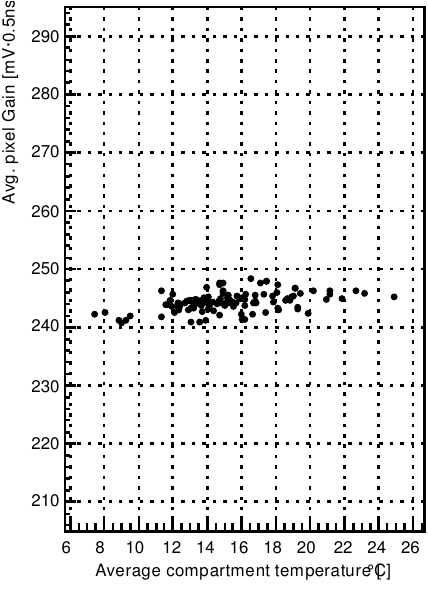}\hfill
\includegraphics[width=0.67\textwidth]{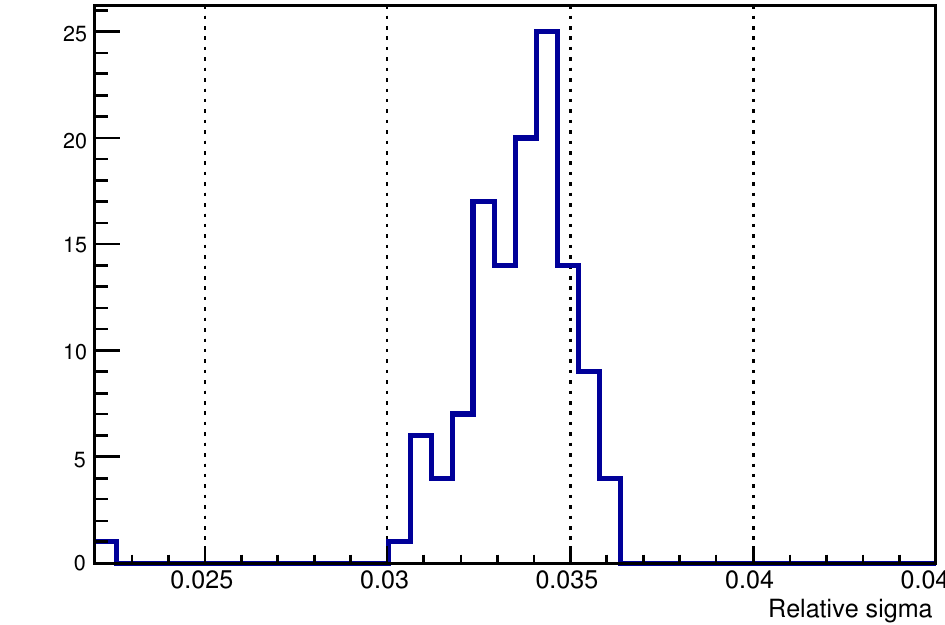}\hfill
\caption{The left figure shows the average gain extracted from 110 measurements
in arbitrary units versus the corresponding average sensor temperature
in degree Celsius. A small remaining dependence is visible, most probably due
to a non-ideal temperature coefficient. The spread of the distribution is
less than 1\%. The right figure shows the distribution of the
corresponding variance relative to the average. The typical spread of
gain values in the camera is 3.4\% on average.} 
\label{fig:gain} 
\end{figure*}

In total, 110 such distributions have been measured during two months with
a temperature range between 6\,\textdegree{}C and 24\,\textdegree{}C
with several measurements each night. Figure~\ref{fig:gain} shows the 
average gain of all pixels versus the average temperature of the 28
temperature sensors. Although the variance of the distribution is less
than 1\%, still a remaining dependence on temperature is visible. It
can be attributed to a non-ideal temperature coefficient. Since the
remaining temperature difference is in the order of less than 1\,mV a
large statistics is needed to reach that precision. A camera with more than 1400
channels is ideally suited for that. The shown data has been used to
apply a correction on the coefficient. Currently, new data is recorded
to investigate the expected improvement. In addition to the average
dependency on temperature, the gain spread in the camera can be
investigated. Figure~\ref{fig:gain} shows the distribution of variances
of the measurements for the gain distributions. It is
independent of the temperature and has an average value of 3.4\%.

For all those measurements, typical currents per sensor are in the
order of 1\,\(\mu\)A or below, which is close to the resolution of the 
current readout and the corresponding voltage drop is within the
resolution of the voltage setting. If the lid is opened, this does not
apply anymore and a correction for the induced voltage drop has to
be introduced.

\subsection{Current dependency}

\begin{figure*}[!htb]
\centering
\includegraphics[width=0.49\textwidth]{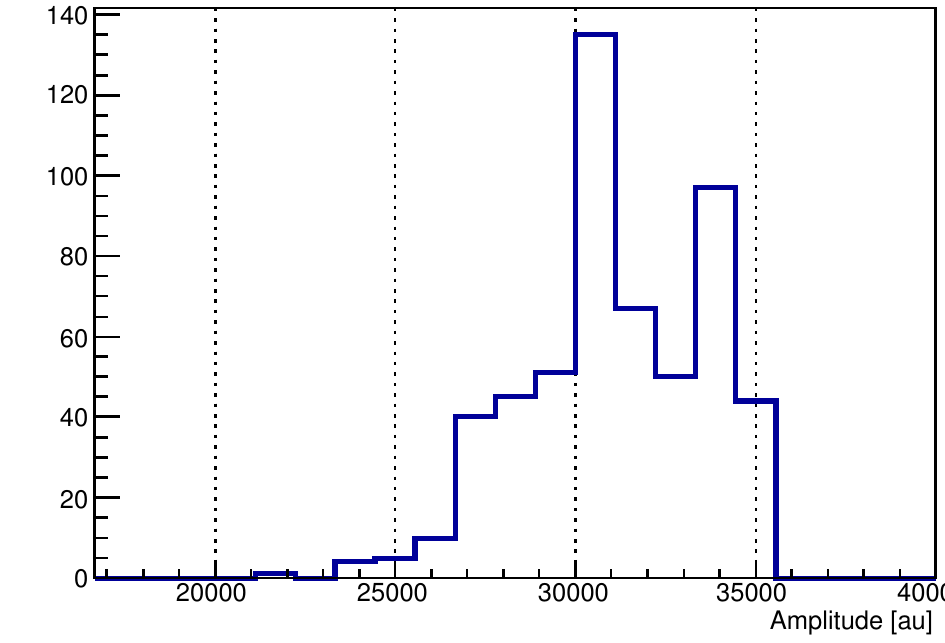}\hfill
\includegraphics[width=0.49\textwidth]{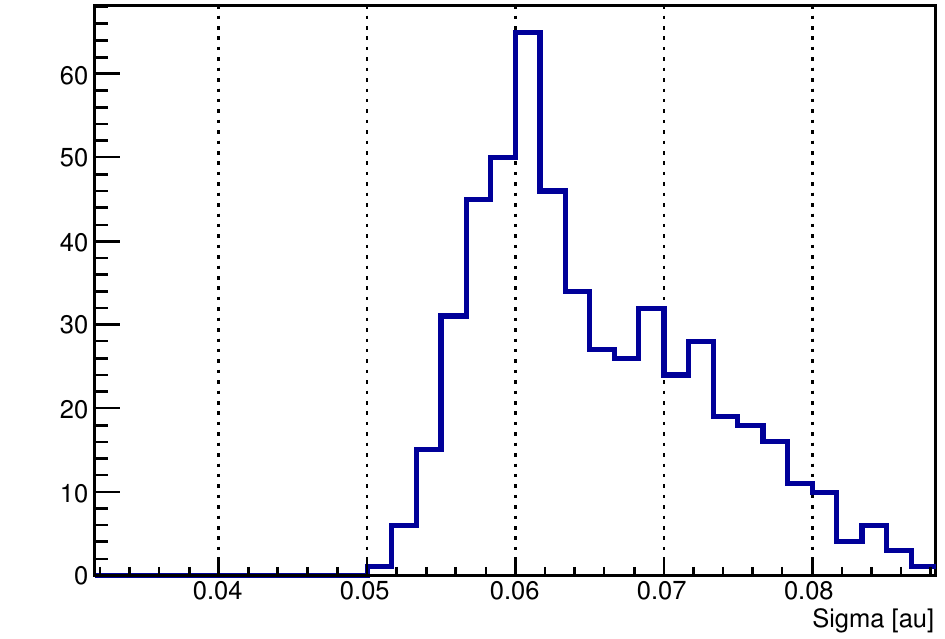}\hfill
\caption{The left figure shows the average amplitude in arbitrary units
extracted from several light-pulser runs, the right figure the distribution
of the spread in the camera relative to the average. The variance of the average
is below 10\%, the typical spread in the camera is around 6\%. This includes
the inhomogeneity of the light yield from the light pulser, the fluctuation
in its light yield and a small remaining temperature dependance of its light yield.} 
\label{fig:lp} 
\end{figure*}

If the camera lid is open, the camera is exposed to the photon flux
from the diffuse night-sky background light resulting in typical 
currents per sensor in the order of 5\,\(\mu\)A to 10\,\(\mu\)A during
dark night. At full moon nights this can reach currents of 200\,\(\mu\)A
per sensor or more.

To measure the dependency on background light level expressed as current
per sensor, dedicated runs of light-pulser data are taken about every
20 minutes during the night.
The distribution of the average amplitude seen in the camera and its
variance are shown in Figure~\ref{fig:lp}. The variance of the
distribution on averages is below 10\%, the mean of the distribution on
the variances is roughly 6.5\%. It should be mentioned that the light
distribution of the pulser has a systematic uncertainty of a few percent and a
remaining temperature dependence of its amplitude also in the order of
a few percent is known. 

\subsection{Data taking efficiency and automation}

\begin{figure*}[!htb]
\centering
\includegraphics[width=0.498\textwidth]{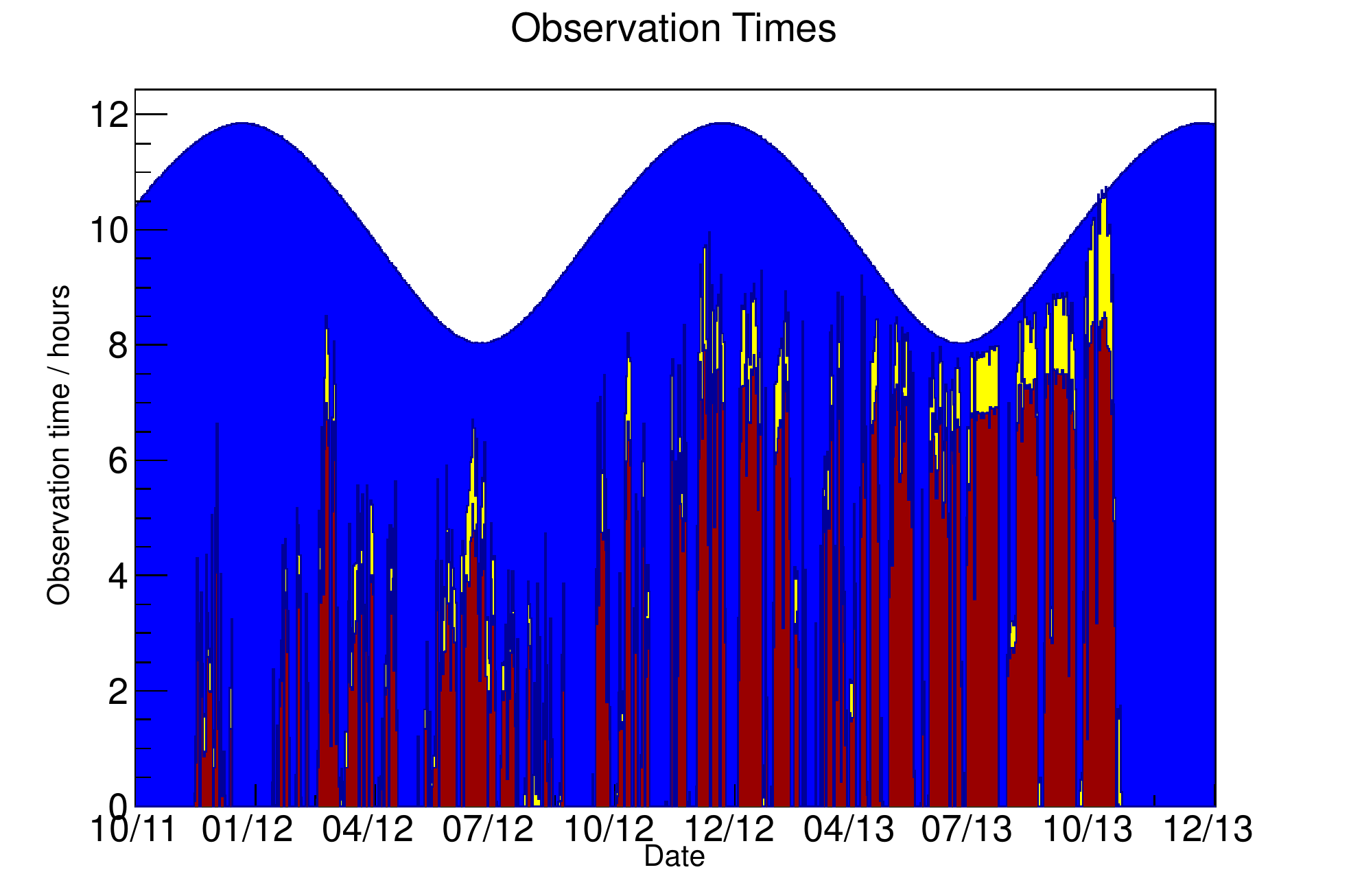}\hfill
\includegraphics[width=0.498\textwidth]{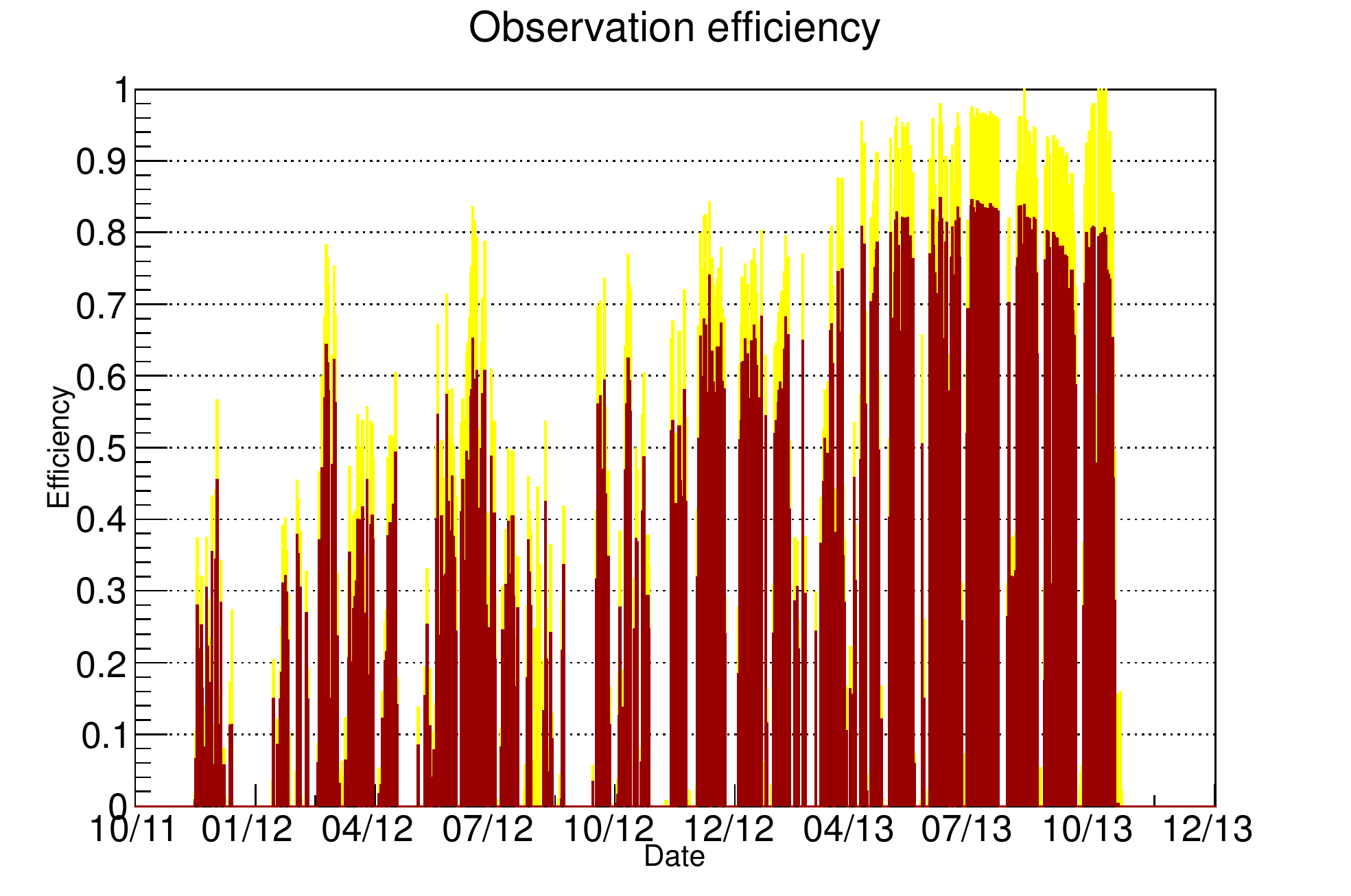}
\caption{The left figure shows the observation total time of data taken
per night (yellow) and the total time of physics triggered data taken (red).
The difference is the amount of calibration data recorded. As a reference, the
total available data taking time is shown (astronomical twilight) in blue. The right
plot shows the corresponding efficiency normalized to the available twilight time.
Roughly since April 2013, data taking is fully automatic which is clearly visible
in the data taking efficiency. The regular gaps originate from full moon days
during which the telescope cannot be operated due to safety reasons. Other
downtime is usually related with bad weather.} 
\label{fig:obs} 
\end{figure*}

To achieve the goal of long-term monitoring of the brightest known TeV
blazars with a high duty cycle, the system has been fully automated.
Although improvements are continuously on-going, the system is now
operated with a pre-defined schedule usually without interruption from
the beginning to the end of the night without manual interaction. The
main reasons for down time are now mainly bad weather conditions and too
bright moon light. Since the light of the moon and thus currents are disproportionately
 hight around full moon, for safety reasons currently an operation limit is
imposed on the currents. In addition, no shift crew is available on the
site of the close-by MAGIC telescope, which prevents remote
operation due to an agreement with the MAGIC collaboration for safety reasons.
Apart from these unavoidable constrains, data taking
efficiency is only limited by the high amount of calibration data still
taken to get a good understanding of the system in all possible
conditions. The least possible amount of time is lost due to the
configuration of the system at the beginning of each run or
re-pointing of the telescope. Starting a run takes about one second,
while re-positioning takes a few seconds for the movement and a few
seconds for the voltage to become stable again after that. In total,
usually less than 40\,s are lost for each 20\,min block of data-taking
which corresponds to an efficiency of about 97\% (including
calibration data). Figure~\ref{fig:obs} shows the total data taking per
night, the time of astronomical twilight. Super-imposed is the total time
the system operated and the time physics data was recorded. The second
plot shows the corresponding efficiencies. It can be seen that since a
few months, the operation efficiency during the night usually exceeds
95\% and the data taking efficiency for physics triggered data exceeds 80\%.
Data taking can be monitored online at http://www.fact-project.org/smartfact.
In addition, work to operate the telescope fully robotically are in-going,
c.f.~\cite{bib:robotic}. 

\subsection{Monitoring}

Due to automation, the resulting high data taking efficiency results in
a dense and complete sampling of the monitored sources. In
\cite{bib:physics} the excess rates as measured from the blazars Markarian 421
and Markarian 501 between May 2012 and July 2013 are shown. First results were
presented in~\cite{bib:Gamma2012}. The
typical observation time is in the order of a few hours per night.
Observations are typically carried out up to a moon disk of 85\% to
90\%. In total 167 hours of data have been collected for Markarian 421
and 242 hours for Markarian 501. Since the data taking procedure was optimized and
data taking is fully automatic, sampling density has significantly improved. The results of
the analysis are immediately publicly available online at
http://www.fact-project.org/monitoring, typically with a delay not less
than 20\,min. For data taking and online analysis in total four to five
cores are loaded on average in two PCs.

\section{Conclusions}

The First G-APD Cherenkov telescope has proven the possibility to apply
Geiger-mode avalanche photo diodes in Cherenkov astronomy for photo
detection. Their gain has been shown to be independent of the
temperature within 1\% in a temperature range between 0\,\textdegree{}C
to 30\,\textdegree{}C and independent from the night-sky background
light level within at least 6.5\%. The gain distribution of the 1440 pixels is not
wider than 3.4\% with a closed lid and 10\% with open lid, which
includes the inhomogeneity of the light-pulser's light distribution
used for the measurement. These stability has been achieved independent of any
external calibration device. The external light-pulser has only be used
to measure the G-APD properties. Currently, more calibration
measurements are taken to improve these results further.

With a high level of automation, the fully remote controlled telescope,
achieves a data taking efficiency of typical 85\% and more than 95\% if
calibration measurement are included. Configuration and re-positioning
times are less than 5\% in total.

Recently, it was proven that observations even close to 
the full moon are possible, c.f.~\cite{bib:moon}.

Monitoring of the two brightest known blazars is carried out
successfully since two years and the excess rates of the past
1.5 years have been present including a couple of minor and major
flares. The results from an online analysis are available online
immediately after data taking.

\vfill

%



%

\end{document}